\documentstyle[12pt]{article}

\def\bc{\begin{center}}
\def\ec{\end{center}}
\def\be{\begin{equation}}
\def\ee{\end{equation}}
\def\myappendix{\par
 \setcounter{section}{0}
 \setcounter{subsection}{0}
 \setcounter{equation}{0}
 \setcounter{table}{0}
 \def\appendixname{Appendix}
 \def\appesection{\setcounter{equation}{0}\section}
 \def\@thesection{\Alph{section}}
 \def\thesection{\appendixname\hskip 1.10ex\Alph{section}}
 \def\thesubsection{\@thesection.\arabic{subsection}}
 \def\theequation{\@thesection.\arabic{equation}}
 \def\thetable{\@thesection.\arabic{table}}}
\newcommand{\np}{Nucl. Phys. }
\newcommand{\pl}{Phys. Lett. }

\newcommand{\nn}{\nonumber}

\newcommand{\beq}{\begin{equation}}
\newcommand{\eeq}{\end{equation}}
\newcommand{\beqn}{\begin{eqnarray}}
\newcommand{\eeqn}{\end{eqnarray}}

\def\vdir{v\kern-7.8pt\Big{/}}
\def\pdir{p\kern-7.8pt\Big{/}}

\begin{document}
\pagestyle{empty} 
\vspace{-0.6in}
\begin{flushright}
CERN-TH/96-31\\
LPTHE 96-09\\
ROME prep. 96/1131 \\
ROM2F/96/06
\end{flushright}
\vskip 0.2 cm
\centerline{\large{\bf{TWO-BODY NON-LEPTONIC DECAYS ON THE LATTICE}}}
\vskip 0.2cm
\centerline{\bf{M. Ciuchini$^{1,a}$, E. Franco$^{2,b}$
G. Martinelli$^{3}$ and L. Silvestrini$^{4}$}}
\centerline{$^1$ Theory Division, CERN, 1211 Geneva 23, Switzerland.}
\centerline{$^{2}$ LPTHE, Univ. de Paris XI, 91405 Orsay, France$^{\rm *}$.}
\centerline{$^3$ Dip. di Fisica, Univ. ``La Sapienza"  and}
\centerline{INFN, Sezione di Roma, P.le A. Moro, I-00185 Rome, Italy.}
\centerline{$^4$ Dip. di Fisica, Univ. di Roma ``Tor Vergata''
and INFN,  Sezione di Roma II,}
\centerline{Via della Ricerca Scientifica 1, I-00133 Rome, Italy.}
\abstract{We show that, under reasonable hypotheses,
it is possible to study two-body non-leptonic weak decays 
in numerical simulations 
of lattice QCD.  By assuming that  
final-state interactions are  dominated by the nearby resonances and 
that the couplings of the resonances to the final particles
are smooth functions of the external momenta,
it is possible indeed to overcome the difficulties imposed by the Maiani-Testa
no-go theorem and to extract the weak decay  amplitudes, including their
phases.
Under the same assumptions, results can be obtained
also for time-like form factors and quasi-elastic 
processes.}\vskip 3cm
\noindent$^a$ On leave of absence  from 
INFN, Sezione Sanit\`a, V.le Regina Elena 299,
00161 Rome, Italy. \\
$^b$ On leave of absence  from INFN, Sezione di Roma I,
Dip. di Fisica, Universit\`a
degli Studi ``La Sapienza", Rome, Italy. \\
$^*$ Laboratoire associ\'e au 
 Centre National de la Recherche Scientifique.

\vfill\eject
\pagestyle{empty}\clearpage
\setcounter{page}{1}
\pagestyle{plain}
\newpage
\pagestyle{plain} \setcounter{page}{1}
 
\section*{Introduction} \label{intro}
Exclusive non-leptonic decays are a fundamental source of information on
quark 
weak inte\-ractions and on the strong interaction
dynamics. Unfortunately,
a theoretical description of exclusive decays based on the fundamental theory
is not  possible yet. Over the years, several me\-thods have been introduced
 to estimate
the relevant matrix elements: vacuum saturation, bag models,
quark models,  QCD sum rules,
$1/N_c$ expansion, chiral Lagrangians, factorization, etc. 
None of these approaches is
fully   satisfactory. In the first three cases, the methods are in no
way a systematic expansion of the fundamental theory in a small parameter.
Thus the result of some calculation
can be either correct or wrong, but it is impossible
to reduce the uncertainty of the theoretical 
prediction  by computing  higher orders in the expansion.
With QCD sum rules it is very difficult to improve the results
in a systematic way because only  few perturbative terms
and matrix elements of higher dimensional operators
can be estimated;  in the $1/N_c$ expansion nobody has succeeded in
computing consistently the matrix elements of the effective
Hamiltonian beyond the lowest order and
  chiral Lagrangians have a rather limited 
domain of application. It may well be true that factorization holds for
weak decays of hadrons containing a heavy quark, although a real proof
that this is the case is still lacking and moreover
there are amplitudes that cannot
be factorized.
\par The lattice approach has been used  
to obtain results based on first principles
  for a wide set of relevant physical quantities such as
the hadron spectrum, the meson decay constants,
the form factors entering in semileptonic and radiative decays, 
the kaon $B$-parameter $B_K$, etc. For  exclusive non-leptonic decays,
however, this method is no better than the above-mentioned approaches, because
of the Maiani--Testa No-Go Theorem  (MTNGT) and the activity
in this field \cite{bss,ab1}
has completely stopped after the publication of ref.~\cite{mt}.
\par In this paper we show that, under quite reasonable  physical hypotheses,
 it is possible to extract predictions
for the relevant matrix elements in numerical simulations
of lattice QCD, in spite of the MTNGT. Our proposal
has a wide domain of applications, ranging from kaon decays,
$K \to \pi \pi$, to charm and beauty decays, $D \to K \pi$, $B \to D \pi$,
$B \to \pi \pi$, etc. If successful, it will also allow a check of several
interesting theoretical ideas as the factorization of the amplitudes 
and the scaling laws 
that can be formulated in the Heavy Quark Effective Theory
(HQET) when the mass of the quark becomes large.
Under the same assumptions, it is possible to try a numerical 
calculation of the time-like form factors, e.g. those relevant
in $e^+ e^- \to \pi^+ \pi^-$ or  $n \, \bar n$, and of those entering
in quasi-elastic electron--nucleon scattering processes,
e.g. $e^- + p \to e^- + \Delta + \pi$.
\par The MTNGT states essentially the following:
\begin{itemize} \item In the calculation
of a two- (many-) body decay amplitude performed 
in the Euclidean space-time, which is 
the only possibility in Monte Carlo simulations,
 there is no distinction between {\it in-} and
{\it out-}states. As a consequence, the matrix elements that one is able
to extract are real numbers resulting from the  average of the two cases, e.g. 
\be \langle \pi \pi \vert {\cal H}_W \vert K \rangle =
\frac{1}{2} \Bigl( \,  _{{\it in}}\langle \pi \pi \vert {\cal H}_W \vert K \rangle
+_{{\it out}}\langle \pi \pi \vert {\cal H}_W \vert K \rangle \Bigr)
\, . \label{eq:inout} \ee
Any trace of the phase due to final-state interactions is then lost. 
This jeopardizes the possibility
of  any realistic prediction for the matrix elements. For example, we know from
the measured $A_{1/2}$ and $A_{3/2}$ amplitudes in $D \to K\pi$ decays
that there is a phase difference of about $80^\circ$.
\item Matrix elements are extracted on the lattice by studying the
time behaviour of appropriate correlation functions at  large time
distances. Maiani and Testa showed  that what can be really isolated
in this limit are the off-shell form factors corresponding to
the final particles at rest, e.g.  $\langle \pi
(\vec p_{\pi}=0) \pi(\vec p_{\pi}=0) \vert {\cal H}_W \vert K \rangle$.
For kaon decays, we can use the chiral theory to extrapolate the
form factor to the physical point\footnote{ 
In the chiral limit final-state interactions are negligible and the problem
discussed before  is  solved. This is thus the only 
situation in which
the matrix element can be evaluated.}.
This is certainly not the case for $D$- and $B$-meson
decays. In the latter case
it is not possible to obtain a realistic prediction for the matrix element.
\end{itemize}
\par We will show that both the  difficulties raised by Maiani and Testa
can be overcome under the hypothesis that final-state interactions are
dominated by nearby resonances, the couplings of which to the final-state
particles satisfy some smoothness condition\footnote{
A different approach, based on the study of the two-particle energy
spectrum in a periodic box can be found in ref.~\cite{luscher}}.
  By varying the spatial momentum
of the initial and final hadrons, it is then possible to extract
the physical matrix elements, including the phase due to the strong-interaction
rescattering of the final states. This allows also the calculation of
the relevant parameters of the resonances. 
\par Two observations are necessary at 
this point. \par The assumption that final-state interactions are dominated by
resonances is not new and has been successfully applied to phenomenological
studies of charmed meson decays, see for example ref.~\cite{lusi}
(together with the factorization hypothesis that we need not assume).
\par At this stage, we do not know if the procedure  that we are proposing
can lead to useful results  in practice. It remains
to be seen whether, with a reasonable size of the lattice 
and number of gauge field
configurations, it is possible to extract the matrix
elements with a satisfactory accuracy.  A feasibility study 
is currently under way on the APE machine. 
\section*{The three-point Euclidean correlation function}
\label{sec:tpf}
Following ref.~\cite{mt}, we first examine the Euclidean three-point function
$G(t_1,t_2,\vec q,-\vec q)$ 
\beq G(t_1,t_2,\vec q,-\vec q)\equiv 
\langle 0 \vert  T \left[ \Pi_{\vec q}(t_1) \Pi_{-\vec q}(t_2) H(0)
\right]  \vert 0 \rangle=
\langle 0 \vert \Pi_{\vec q}(t_1) \Pi_{-\vec q}(t_2) H(0)
\vert 0 \rangle\, , \label{eq:3pf}
\eeq when $t_1>t_2>0$.
In eq.~(\ref{eq:3pf}),  $\Pi_{\vec q}(t)$ is an interpolating field
of the final-state particle (denoted as  ``pion"
in the following) with a fixed spatial momentum
\be \Pi_{\vec q}(t) = \int \,  d^3x \,  e^{- i\vec q \cdot \vec x}
\Pi (\vec x , t)\, ; \ee
$H(0)=H(\vec x=0, t=0)$ is any local operator that couples to the
two pions in the final state; $ T \left[ \dots \right]$ represents
the $T$-product of the fields and   the vacuum
expectation value corresponds, in a numerical simulation, to the
average over the gauge field configurations. \par
When $t_1 \to \infty$
\beqn &\,&G(t_1,t_2,\vec q,-\vec q)
 \to \sum_n \langle 0\vert  \Pi_{\vec q}(t_1) 
\vert n \rangle \langle n \vert \Pi_{-\vec q}(t_2) H(0) \vert 0 \rangle
 \sim \frac{\sqrt{Z_\Pi}}{2 E_{\vec q}} e^{-E_{\vec q}t_1} G_3(t_2)
\, ,\eeqn
where
\be E_{\vec q}=\sqrt{M_\pi^2+{\vec q}^2}\,,\quad 
\sqrt{Z_\Pi} =\langle 0 \vert \Pi(0) \vert \vec q \rangle \ee
and 
\beq G_3(t)=\langle \vec q\, \vert \Pi_{-\vec q}(t) H(0)\vert 0 \rangle \eeq
for $t>0$. 
\par Inserting a complete set of {\it out}-states, we can write
\beqn&\,& \langle \vec q \, \vert \Pi_{-\vec q}(t) H(0)\vert 0 \rangle 
=\sum_n (2 \pi)^3
\delta^3(P_n)
\langle \vec q \, \vert e^{\hat H t} \Pi(0)e^{-\hat H t}
\vert n \rangle_{out} \,_{out}\langle n \vert H(0) \vert 0 \rangle =
\nn \\ &\,&\sum_n (2 \pi)^3
\delta^3(P_n)
 \langle \vec q \, \vert\Pi(0)
\vert n \rangle_{out} \,_{out}\langle n \vert H(0) \vert 0 \rangle
e^{-(E_n-E_{\vec q}) t} = \nn \\ &\,&
\frac{\sqrt{Z_\Pi}}{2E_{\vec q}} e^{-E_{\vec q} t} \times
\,_{out}\langle \vec q, -\vec q \vert H(0) \vert 0 \rangle + \nn \\ &\,&
\left[\sum_n (2 \pi)^3 \delta^3(P_n)
 \langle \vec q\, \vert\Pi(0) \vert n  \rangle_{out}
\,_{out}\langle n \vert H(0) \vert 0 \rangle
e^{-(E_n-E_{\vec q}) t}\right]_{\mbox{\tiny connected}}\, ,
\label{starting}\eeqn
where  the  term 
proportional to $_{out}\langle \vec q,- \vec q \vert H(0) \vert 0\rangle$
on the r.h.s. is the disconnected
contribution. 
We now make the following assumptions:
\begin{enumerate}
\item The only possible eigenstates of the Hamiltonian
are $n$-pion states.
\item The interaction between  pions is dominated
by a narrow resonance (denoted as  $\sigma$ in the following) exchanged 
in the $s$-channel. 
\item The coupling of $\sigma$ to the pions
is a smooth function of the external momenta.
\end{enumerate}
Under hypotheses 1. and 2., the only possible intermediate states
are two-pion states. We thus obtain 
\beqn &\,&
G_3(t)= \sum_n (2 \pi)^3
\delta^3(P_n)
 \langle \vec q \, \vert\Pi(0)
\vert n \rangle_{out} \,_{out}\langle n \vert H(0) \vert 0 \rangle
e^{-(E_n-E_{\vec q}) t} =\nn \\ &\,& 
\frac{\sqrt{Z_\Pi}}{2E_{\vec q}} e^{-E_{\vec q} t} \times
 \frac{g(s_q)}{M^2_\sigma-s_q-i X(s_q)} +
\nn \\ &\,& \left[\sum_n (2 \pi)^3 \delta^3(P_n)
 \langle \vec q\, \vert\Pi(0) \vert n  \rangle_{out}
\,_{out}\langle n \vert H(0) \vert 0 \rangle
e^{-(E_n-E_{\vec q}) t}\right]_{\mbox{\tiny connected}}\,.
\label{master}\eeqn
In eq.~(\ref{master}) $\sqrt{s_q}=2E_{\vec q}$ ; $\vert n \rangle_{out}=
\vert \vec k, - \vec k \rangle_{out}$;
$\sqrt{s}=E=2 E_{\vec k}$; the sum over
the intermediate states is given by\footnote{A further factor $1/2$
in the case of identical particles is understood.}
\beqn &\,&
\sum_n (2 \pi)^3 \delta^3(P_n) =
\int \frac{d^3k_1}{(2\pi)^3 2 E_{\vec k_1}}
\frac{d^3k_2}{(2\pi)^3 2 E_{\vec k_2}} (2 \pi)^3\delta^3(\vec k_1+
\vec k_2)=\nn
\\ &\,&\int \frac{dE}{2\pi}
\int \frac{d^3k_1}{(2\pi)^3 2 E_{\vec k_1}}
\frac{d^3k_2}{(2\pi)^3 2 E_{\vec k_2}} (2 \pi)^4\delta^3(\vec k_1+\vec k_2)
\delta(E_{\vec k_1}+E_{\vec k_2}-E)\,  \eeqn
and we have used
\beq  _{out}\langle \vec q,-\vec q \vert H(0) \vert 0 \rangle =\frac{g(s_q)}{
M^2_\sigma-s_q-i X(s_q)}\vert_{s_q=p_H^2}\, . \label{uno}\eeq
\par We now define
\beqn 
\langle \vec q \,\vert\Pi(0) \vert\vec k,-\vec k  \rangle_{out}
&=& 
\left[ \frac{2\sqrt{Z_\Pi}}{(E+2 E_{\vec q}-i \epsilon)
(-E+2 E_{\vec q}-i \epsilon)}\frac{\vert V(s)
 \vert^2}{M^2_\sigma-s-i X(s)}
\right]^* \nn \\ &=&\frac{2\sqrt{Z_\Pi}}
{(E+2 E_{\vec q} +i \epsilon)
(-E+2 E_{\vec q}+
i \epsilon)}\frac{\vert V(s) \vert^2}{M^2_\sigma-s+i X(s)}
\label{defi}\eeqn
In eq.~(\ref{defi}) we have introduced the $\sigma$--$\pi$--$\pi$ coupling
$V(s)$. As stated before, we assume here that the coupling is such a smooth 
function of the external momenta that we can use the same  
``physical"  coupling also for the  off-shell pion, which is
annihilated  by $\Pi(0)$
in the   matrix element 
$\langle \vec q \,\vert\Pi(0) \vert \vec k,-\vec k  \rangle_{out}$. 
In
general we should write
\beqn  _{out}\langle \vec k,- \vec k \vert \Pi(0)
 \vert \vec q \rangle &=&\sqrt{Z_\Pi}\times \left[ \frac{2}
{(E+2 E_{\vec q}-i \epsilon)
(-E+2 E_{\vec q}-i \epsilon)}\right]\times \nn \\
\frac{\vert V(s) \vert^2}{M^2_\sigma-s-i
X(s)} &\times& {\cal F}\left(1-\frac{E}{2E_{\vec q}}\right) \, ,
\label{defig}\eeqn
with the condition that the modulating factor ${\cal F}$
of the off-shellness $1-E/2E_{\vec q}$ satisfies the condition ${\cal F}(0)=1$
(see below). The factor in parenthesis is (up to a factor
$1/2 E_{\vec q}$) the propagator of two non-interacting
pions \cite{cm}.
\par
With the definition in eq.~(\ref{defi}),
 the quantity in square brackets satisfies
\beqn &\,&  \lim_{E=\sqrt{s} \to 2 E_{\vec q}} (-p^2+M_\pi^2)
\left[\frac{2\sqrt{Z_\Pi}}{(E+2 E_{\vec q}-i \epsilon)
(-E+2 E_{\vec q}-i \epsilon)}\frac{\vert V(s) \vert^2}{M^2_\sigma-s-i X(s)}
\right] \to \nn \\ &\,& 
\sqrt{Z_\Pi} \frac{\vert V(s_q) \vert^2}{M^2_\sigma-s_q-i 
X(s_q)}
= \sqrt{Z_\Pi} {\cal A}\Bigl(\pi(\vec q)+\pi(- \vec q) \to 
n=\pi(\vec k)+\pi(-\vec k)\Bigr)\vert_{\vert \vec k \vert=\vert \vec q \vert}\,
,  \label{onsh}\eeqn
which is the usual LSZ-reduction formula for on-shell particles, cf.
 eq.~(15) of ref.~\cite{mt}. In eq.~(\ref{onsh}), $p^2=E^2+M^2_\pi-2E
\, E_{\vec q}$ is the squared four-momentum of the off-shell pion.
Combining eqs.~(\ref{master}) and (\ref{defi})
and using the identity
\beq \frac{1}{E-2E_{\vec q}-i\epsilon}= {\cal P}\left[
\frac{1}{E-2E_{\vec q}}\right]+i\pi\delta(E-2E_{\vec q})\,
,  \eeq
 we obtain
\beqn G_3(t)&=&\frac{\sqrt{Z_\Pi}}{2 E_{\vec q}} e^{E_{\vec q}t}
\times  
\left[\frac{g(s_q)}{M^2_\sigma-s_q-i X(s_q)} e^{-2 E_{\vec q}t}
\right .\nn \\ &\,&\left.
+ (4 E_{\vec q})  \int \frac{dE}{\pi}
 \frac{ g(s)}{(E+2 E_{\vec q}+i \epsilon)
(-E+2 E_{\vec q}+i \epsilon)} \frac{X(s)}{(M^2_\sigma-s)^2+ X^2(s)}
e^{-Et} \right]\nn \\ &=&
\frac{\sqrt{Z_\Pi}}{2 E_{\vec q}} e^{E_{\vec q}t}
\times 
\left\{\left(\frac{g(s_q)}{M^2_\sigma-s_q-i X(s_q)}
- \frac{i g(s_q)X(s_q)}{(M^2_\sigma-s_q)^2+ X^2(s_q)}\right) e^{-2 E_{\vec q}t}
\right . \nn \\ &\,&\left . 
+ (4 E_{\vec q}) {\cal P}\left[\frac{1}{\pi} \int_{2 M_\pi}^{+\infty}
dE e^{-E t} \frac{g(s)\rho(s)}{4E^2_{\vec q}-E^2}\right]\right\}\, ,
\label{mresult}\eeqn 
where, in the integrand of the r.h.s.,
 we have denoted as  $E^2$ or  $s$  the same quantity, $s=E^2$. This 
will facilitate the comparison of the above formulae with 
the more general case of a two-pion
final state with total momentum different from zero, cf. eqs.~(\ref{exampleq})
and (\ref{deff}) below.
In eq.~(\ref{mresult}) we have used the relation $X(s)= M_\sigma \Gamma(s)$,
where $\Gamma(s)$ is the ``width" of the $\sigma$-meson defined as
\beqn \Gamma(s) &=&\frac{1}{2 M_\sigma} \Bigl[
\int \frac{d^3k_1}{(2\pi)^3 2 E_{\vec k_1}}
\frac{d^3k_2}{(2\pi)^3 2 E_{\vec k_2}} (2\pi)^4 \delta^3(\vec k_1+\vec k_2)
\delta(E_{\vec k_1}+E_{\vec k_2}-\sqrt{s}) \vert V(s)\vert^2 \Bigr] \nn \\
&=& \frac{ \vert V(s)\vert^2 }{8 \pi M_\sigma} 
\sqrt{\frac{(s-4 M_\pi^2)}{4 s}}
\, \theta(s-4 M_\pi^2)\, .\label{gamma} \eeqn
\par The spectral function $\rho(E^2)$ is given by
\beqn \rho(E^2)= \frac{X(E^2)}{(M_\sigma^2-E^2)^2+X(E^2)^2} \, , \nn \eeqn
which becomes
\beqn
\rho(E^2)= \frac{1}{2 M_\sigma}\frac{\Gamma(E^2)/2}{(M_\sigma-E)^2+\left(\Gamma(E^2)
/2\right)^2} \eeqn
for a narrow resonance. We also have  
\be \rho(E^2) \to \frac{\pi}{2 M_\sigma} \, \delta ( M_\sigma-E) \ee
as $\Gamma \to 0$.
\par We then establish the relation
\beqn &\,& \frac{g(s_q)}{M^2_\sigma-s_q-i X(s_q)}-
\frac{i g(s_q)X(s_q)}{(M^2_\sigma-s_q)^2+ X^2(s_q)}=\nn \\&\,&
\frac{g(s_q)}{M^2_\sigma-s_q-i X(s_q)}+
\frac{1}{2}\left(\frac{g(s_q)}{M^2_\sigma-s_q+i X(s_q)}-
\frac{g(s_q)}{M^2_\sigma-s_q-i X(s_q)}\right)=\nn \\ &\,&
_{out}\langle \vec q,-\vec q \vert H(0) \vert 0\rangle
+\frac{1}{2}\Bigl(\,
_{in}\langle \vec q,-\vec q \vert H(0) \vert 0\rangle
-\,
_{out}\langle \vec q,-\vec q \vert H(0) \vert 0\rangle \Bigr)\, , \eeqn
in agreement with eq.~(21) of ref.~\cite{mt}. The last term in 
eq.~(\ref{mresult}) corresponds to the last term
of eq.~(23) of ref.~\cite{mt}.
The definitions  of 
$_{out}\langle \vec q,-\vec q \vert H(0) \vert 0\rangle$
and  $_{in}\langle \vec q,-\vec q \vert H(0) \vert 0\rangle$
are  in agreement with the standard relation between the strong rescattering 
phase and the parameters of the resonance $\sigma$
\beqn _{out}\langle \vec q,-\vec q \vert \vec q,-\vec q
\rangle_{in}=  e^{2 i \delta(s)}&=&\frac{M^2_\sigma-s +i X(s)}{M^2_\sigma-s -i X(s)}\, ,
\label{phase} \eeqn
where the phase $\delta(s)$ is defined up to an ambiguity
of $\pi \equiv 180^\circ$. We choose $0 \le \delta(s) \le \pi$.
\par In the case of non-interacting pions, 
$_{out}\langle \vec q,-\vec q \vert H(0) \vert 0\rangle=\,  _{in}\langle \vec
q,-\vec q \vert  H(0) \vert 0\rangle=\sqrt{Z_\sigma}$, and one finds
\beq G(t_1,t_2,\vec q, -\vec q)=
\left(\frac{\sqrt{Z_\Pi}}{2 E_{\vec q}} \right)^2  e^{-E_{\vec q}(t_1+t_2)}
\sqrt{Z_\sigma}\, .\label{expli}\eeq
Equation (\ref{expli}) has a very simple interpretation. The two
non-interacting particles
are created in the origin $t=0$ with amplitude $\sqrt{Z_\sigma}$ and propagate
from the origin to $t_2$; from $t_2$
to $t_1$ only  one particle state propagates.
\par In the limit of a zero-width resonance ($X(s) \to 0$) we obtain
\beq G(t_1,t_2,\vec q, -\vec q)=
\left(\frac{\sqrt{Z_\Pi}}{2 E_{\vec q}} \right)^2 
\frac{e^{-E_{\vec q}(t_1+t_2)}}{M^2_\sigma-s_q} \left[g(s_q)-g(M^2_\sigma)
\left(\frac{2 E_{\vec q}}{M_\sigma}\right) e^{-(M_\sigma-2 E_{\vec q})t_2}
\right]  \, , \eeq
which is in agreement with the result, obtained for a narrow
resonance,  given in  eq.~(3.3)
of ref.~\cite{cm}.  We notice that eq.~(3.3) of ref.~\cite{cm} is really
 valid only at small time-distances. According to  the MTNGT instead, 
for any non-zero value of the width,
 the correlation function at large  time-distances
is always dominated by the state with the two pions at rest.
\par For further use, it is convenient to introduce the quantity
$ R^{H}(t_2, \vec q)$ defined 
from the relation 
\be G(t_1,t_2,\vec q, -\vec q)=
 \left(\frac{\sqrt{Z_\Pi}}{2E_{\vec q}}\right)^2
 e^{-E_{\vec q}(t_1+t_2)} R^{H}(t_2, \vec q)
\, , \label{cambiata} \ee
where the label $H$ in $ R^{H}(t_2, \vec q)$ denotes the local operator
used to create the two pions.
\par It is straightforward at this point to derive the expression
that enters in a two-body  non-leptonic decay of a meson
(which we will denote as $D$ in the following). The starting
point is the four-point correlation function
\beqn G(t_1,t_2,t_D,\vec q,-\vec q,\vec p_D=0)&=& \lim_{t_D \to
-\infty,t_1\to+\infty} \langle \Pi_{\vec q}(t_1)
 \Pi_{-\vec q}(t_2) {\cal H_W}(0)
 D^\dagger_{\vec p_D=0}(t_D) \rangle
\nn \\
&=&\left(\frac{\sqrt{Z_\Pi}}{2E_{\vec q}}\right)^2 e^{-E_{\vec q}(t_1+t_2)}
 R^{{\cal H}_W}(t_2, \vec q) \frac{\sqrt{Z_D}}{2 M_D} e^{-M_D \vert t_D \vert}
\, ,\label{Wdecay}\eeqn
with 
\beqn &\,&  R^{{\cal H}_W}(t_2, \vec q) =
\frac{g^W(s_q)}{M^2_\sigma-s_q-i X(s_q)}+ 
\frac{1}{2}\left(\frac{g^W(s_q)}{M^2_\sigma-s_q+i X(s_q)}-
\frac{g^W(s_q)}{M^2_\sigma-s_q-i X(s_q)}\right) + \nn \\ &\,&
(4 E_{\vec q}) {\cal P} \left[
\int_{2 M_\pi}^{+\infty} 
\frac{dE}{\pi}\frac{e^{-(E-2E_{\vec q})t_2}}{4E_{\vec q}^2-E^2}
\frac{1}{2 i}\left(\frac{g^W(s)}{M^2_\sigma-s-i X(s)}-
\frac{g^W(s)}{M^2_\sigma-s+i X(s)}\right) \right]\, , \eeqn
and 
\beq \,_{out}\langle \vec k,- \vec k \vert {\cal H}_W(0)
 \vert D(\vec p_D=0)\rangle =\frac{g^W(s)}{M^2_\sigma-s-i X(s)}\, ,\eeq
where ${\cal H}_W$ is the weak Hamiltonian. We have not considered complications
coming from the presence of (axial-)vector particles and different flavours.
The formulae reported in this section can easily be extended to these cases. 
\section*{The Watson theorem and physical interpretation of the results}
Let us start from 
\beqn && G(t_1,t_2,\vec q,-\vec q) =
\left(\frac{\sqrt{Z_\Pi}}{2E_{\vec q}}\right)^2  
e^{-E_{\vec q}(t_1+t_2)} \times
\left\{\frac{g(s_q)}{M^2_\sigma-s_q-i X(s_q)}+ \right. \nn \\
&\,&\frac{1}{2}\left(\frac{g(s_q)}{M^2_\sigma-s_q+i X(s_q)}-
\frac{g(s_q)}{M^2_\sigma-s_q-i X(s_q)}\right) +  \\
&\,& \left. (4 E_{\vec q}) {\cal P} \left[
\int_{2 M_\pi}^{+\infty} 
\frac{dE}{\pi}e^{-(E-2E_{\vec q})t_2} \frac{1}{4E_{\vec q}^2-E^2}
\frac{1}{2 i}\left(\frac{g(s)}{M^2_\sigma-s-i X(s)}-
\frac{g(s)}{M^2_\sigma-s+i X(s)}\right) \right]\right\}
\nn \, , \eeqn
and introduce the following relation
\beq \,_{out}\langle \vec k,-\vec k \vert H(0)
 \vert 0 \rangle =\frac{g(s)}{M^2_\sigma-s-i X(s)}= e^{i\delta(s)} A(s)\, .
\label{out}\eeq
The Watson theorem ensures us that (in the absence of CP violation)
$A(s)$ is a real quantity \cite{elementarewatson}. It follows that
\beq g(s)= \sqrt{(M^2_\sigma-s+i X(s))(M^2_\sigma-s-i X(s))}\times  A(s)\eeq 
is  also a real quantity.
 Equation (\ref{phase}) implies 
\beq \,_{in}\langle \vec k,- \vec k \vert H(0)
 \vert 0 \rangle =\frac{g(s)}{M^2_\sigma-s+i X(s)}=e^{-i\delta(s)} A(s)\, .
\label{in} \eeq
We may then write 
\beqn  G(t_1,t_2,\vec q,-\vec q) =
\left(\frac{\sqrt{Z_\Pi}}{2E_{\vec q}}\right)^2  
e^{-E_{\vec q}(t_1+t_2)} \times\label{adelta}  \eeqn
\beqn \left\{A(s_q) \cos\delta(s_q) +
 (4 E_{\vec q}) {\cal P} \left[
\int_{2 M_\pi}^{+\infty} 
\frac{dE}{\pi}e^{-(E-2E_{\vec q})t_2} \frac{1}{4E_{\vec q}^2-E^2}
A(s)\sin\delta(s) \right]\right\}\nn
 \, , \eeqn
with 
\beqn  \cos \delta(s)&=&\frac{M^2_\sigma-s}{\sqrt{(M^2_\sigma-s)^2+X(s)^2}} \\
 \sin \delta(s)&=&\frac{X(s)}{\sqrt{(M^2_\sigma-s)^2+X(s)^2}} \, .\eeqn
\par Notice that at threshold $\delta(s)\vert_{s=4 M_\pi^2}=0$, cf.
eq.~(\ref{gamma}).
\par Following ref.~\cite{mt}, we can compute the leading behaviour of
$G(t_1,t_2,\vec q=0,-\vec q=0)$ for large $t_2$ and express the result
in terms of the parameters of the resonance. We obtain
\be G(t_1,t_2,\vec q=0,-\vec q=0)=
\frac{Z_\Pi}{4M_\pi^2}e^{-M_\pi(t_1+t_2)}A(4M_\pi^2)\,\left[1-a
	\sqrt{\frac{M_\pi}{\pi t_2}}\right]\, ,
	\label{g002}
\ee
where the scattering length $a$ is given by 
\begin{equation}
	a=\frac{\vert 
	 V(4M_\pi^2)\vert^2}{16\pi M_\pi}\frac{1}
	 {M_\sigma^2-4M_\pi^2}\, ,
	\label{a}
\end{equation}
in agreement with eq.~(10) of ref.~\cite{mt}\footnote{ There is a difference of
a factor of two, due to the fact that in our case the particles are 
distinguishable.}.
\section*{Applications to lattice calculations}
For $D$-decays eq. (\ref{adelta})  becomes
\beqn &\,& G(t_1,t_2,t_D,\vec q,-\vec q,\vec p_D=0) =
\left(\frac{\sqrt{Z_\Pi}}{2E_{\vec q}}\right)^2  
e^{-E_{\vec q}(t_1+t_2)} \times  \nn \\ &\,&
\left\{A^W(s_q) \cos\delta(s_q) +
 (4 E_{\vec q}) {\cal P} \left[
\int_{2 M_\pi}^{+\infty} 
\frac{dE}{\pi}e^{-(E-2E_{\vec q})t_2} \frac{1}{4E_{\vec q}^2-E^2}
A^W(s)\sin\delta(s) \right]\right\}
\nn \\ &\,& \times \frac{\sqrt{Z_D}}{2M_{D}}
e^{-M_{D}\vert t_D\vert} \, , \label{ddelta}\eeqn
where 
$g^W(s)= \sqrt{(M^2_\sigma-s+i X(s))(M^2_\sigma-s-i X(s))}\times  A^W(s)$.
\par For numerical applications, it is convenient
to consider the amputated correlation function given by
the ratio
\beq R^{{\cal H}_W}(t_2,\vec q) =
\frac{G(t_1,t_2,t_D,\vec q,-\vec q,\vec p_D=0)}{S_\Pi(t_1,E_{\vec q})
S_\Pi(t_2,E_{\vec q})S_D(t_D, M_D)}\, ,\eeq
where \beq S_\Pi(t_1,E_{\vec q})=\frac{\sqrt{Z_\Pi}}{2E_{\vec q}}
e^{-E_{\vec q}t_1}\, , \label{ratio}\eeq
and similarly for the other meson propagators;
\beq R^{{\cal H}_W}(t_2,\vec q) =
\left\{A^W(s_q) \cos\delta(s_q) +(4 E_{\vec q}) {\cal P} \left[
\int_{2 M_\pi}^{+\infty} 
\frac{dE}{\pi}e^{-(E-2E_{\vec q})t_2} \frac{1}{4E_{\vec q}^2-E^2}
A^W(s)\sin\delta(s) \right]\right\}\, .\label{example} \eeq
\par Until now,
for simplicity, we had chosen $\vec p_D=0$; the generalization of the  
above formulae to $\vec p_D \neq 0$, corresponding  to
$G(t_1,t_2,t_D,\vec q_1,-\vec q_2,\vec p_D=\vec q_1-\vec q_2)$,
is straightforward:
\beqn R^{{\cal H}_W}(t_2,\vec q_1, -\vec q_2) =
\frac{G(t_1,t_2,t_D,\vec q_1,-\vec q_2,\vec p_D=
\vec q_1-\vec q_2)}{S_\Pi(t_1,E_{\vec q_1})
S_\Pi(t_2,E_{\vec q_2})S_D(t_D, E_{\vec p_D})}= \nn  \eeqn
\beqn 
\left\{A^W(s_T) \cos\delta(s_T) +(2 E_{T}) {\cal P} \left[
\int_{E_{min}}^{+\infty} 
\frac{dE}{\pi}e^{-(E-E_{T})t_2} \frac{1}{E^2_T-E^2}
A^W(s)\sin\delta(s) \right]\right\}\, ,\label{exampleq} \eeqn
where 
\beq E_{T}=E_{\vec q_1}+ E_{\vec q_2}\,; \quad
E_{min}=\sqrt{4M_\pi^2+(\vec q_1-\vec q_2)^2} \, ; \quad
s_T=E^2_T-(\vec q_1-\vec q_2)^2  \label{deff}\eeq
and $ \quad s=E^2-(\vec q_1-\vec q_2)^2 $.
\vskip 0.3 cm
\par We now explain the strategy to extract the physical
information, i.e. the matrix element of ${\cal H}_W$, including the phase,
from $R^{{\cal H}_W}(t_2,\vec q)$. The extension to 
$R^{{\cal H}_W}(t_2,\vec q_1, -\vec q_2)$ is 
straightforward.
\par Let us imagine to make a calculation on a lattice of
volume $L^3 \times T$.
The lattice version of eq.~(\ref{mresult}) is given by\footnote{
 We assume here that the discretized version of the two-body phase space 
is a good approximation of the continuum one. In this
case  we can change the integrals
into sums over discrete values of momenta or energies without problems.
Further technical complications arising from the different density of states
in the continuum and on the lattices currently used in numerical simulations
will be discussed elsewhere.}:
\beq R^{{\cal H}_W}(t_2,\vec q) =
A( s_q) \cos\delta(s_q) +
\left(\frac{4 E_{\vec q}}{\pi}\right) 
{\sum_{E_i}^{}}^\prime\left[ \Delta E_i
 e^{-(E_i- 2 E_{\vec q})t_2} \frac{1}{ 4 E_{\vec q}^2-E_i^2}
A(s)\sin\delta(s) \right]\, , \label{examplel} \eeq
where all the quantities are given in units of the lattice spacing,
 \beqn E_i = \sqrt{s}=
2 E_{\vec k} \quad \mbox{with} \quad 
\vec k\equiv \frac{2 \pi}{L}(n_x,n_y,n_z) \,  ;\eeqn
$s_q$ has been defined before; $n_{x,y,z}=0,1,\dots,L-1$ and 
 $\sum_{E_i}^\prime$ denotes the sum over all the 
values of the energy corresponding to  the momenta $\vec k$ allowed by
the discretization of the space-time on a finite volume, excluding those
corresponding to
$E_i= 2 E_{\vec q}$. Different
combinations of momenta corresponding to the same energy should be included only
once in the sum appearing in eq.~(\ref{examplel}), since
the  factors multiplying $\sum_{E_i}^\prime$  already account for the phase
space integration, at  fixed total energy.
Thus, for example,  $\vert k\vert =2\pi/L$
corresponds to six  possibilities\footnote{ In the following, 
$\vec k \equiv 2 \pi/{L}(n_x,n_y,n_z)$
will be simply denoted by $(1,0,0)$.}, $(\pm 1,0,0)$, $(0,\pm 1,0)$, and
$(0,0,\pm 1)$, but has to be counted as a single term in the
sum. $\Delta E_i= E_{i+1}-E_i$ is the difference of the nearest successive
allowed values of $E_i$ ($E_0 = 2 M_\pi$, $E_1=
2\sqrt{M_\pi^2+ (2 \pi/La)^2}$, $E_2=2\sqrt{M_\pi^2+2 (2 \pi/La)^2}$,
etc.). One can show that the expression in eq.~(\ref{examplel}) tends to
the corresponding continuum one in eq.~(\ref{example}) as $L \to \infty$.
\vskip 0.3 cm
 By varying the  external momenta   we can study
\beq \tan\delta(s_q)=\frac{A_s}{A_c} \quad
\mbox{and}\, \,\,\, \vert A(s_q)\vert=\sqrt{A_c^2+A_s^2} \eeq
as a function of the centre-of-mass energy,
where $A_c=A(s_q) \cos \delta(s_q)$ and $A_s=A(s_q)
 \sin \delta(s_q)$. From the behaviour of $\tan \delta(s_q)$ as a function
of $s_q$,  we  can reconstruct the mass and width of the resonance and
extrapolate the amplitude to the physical point.
\par 
$A_c$ and $A_s$ can be found by studying the dependence of
$R^{{\cal H}_W}(t_2, \vec q)$ 
on $t_2$, at fixed $\vec q$, for $t_1 \gg t_2 \gg 0$.
\par Several  observations are important here:
\begin{itemize} 
\item The range of values of $s_q$, which is available in current
numerical studies, is limited because
the momenta on the lattice are quantized and, in order to avoid discretization 
errors, $\vert q \vert  \ll 1$
($\vert q  \vert a \ll 1$, where $a$ is the lattice spacing, if we
do not work in lattice units).  In order to enlarge
the range of values of $s_q$,  it is convenient to have a $D$-meson
with non-zero momentum, corresponding to $\vec q_1 \neq \vec q_2$.
\item To derive the properties of the resonance, it is not necessary
to use the four-point correlation function. We can instead
study the three-point $H$-$\Pi$-$\Pi$ correlation 
 $G(t_1, t_2, \vec q_1, - \vec q_2)$ as a function of the
momenta.  There are several advantages in doing so.
First of all, an additional information on the parameters
of the resonance  can be exploited 
when analysing the four-point correlation
function. Moreover the signal of
the
$H$-$\Pi$-$\Pi$ correlation is expected to be 
much less noisy than in the four-point case. 
A final advantage is that, by varying the quantum
numbers of the field $H(0)$ we can excite and study different resonances,
for example a vector-like one.
\end{itemize}
\par  We now present an example to show how $R^{{\cal H}_W}(t_2,\vec q)$
would appear as a function of $t_2$ for specific values of 
the mass and width of the resonance.
What happens is the following. In the absence of final-state
interactions, $R^{{\cal H}_W}(t_2, \vec q)$ would display a plateau as
a function of $t_2$. In the presence of the resonance,
the second term in eq.~(\ref{examplel})
increases (decreases) exponentially in $t_2$ for states
with $E_i < 2 E_{\vec q}$ ($E_i >  2 E_{\vec q}$). 
It can easily be seen that only the states
with $E_i<  2 E_{\vec q}$ give an appreciable contribution:
the
last term on the r.h.s. of eq. (\ref{examplel}) gives a dramatic 
effect on the plateau for large values of $t_2$.
This is shown  in fig. 1, where
the following parameters have been chosen:
\begin{enumerate}
\item $L=24$ and $T=64$;
\item $\beta=6.2$, corresponding to $a^{-1}\sim 2.9$ GeV; 
\item a resonance with  a mass $M_\sigma=1.9$ GeV and
a  width $\Gamma(M_\sigma^2)=0.3$ GeV, as was  used for $D$-decays 
in ref.~\cite{lusi};
\item $\Gamma(s)=\Gamma(M_\sigma^2)
\sqrt{[(s-4 M_\pi^2)M_\sigma^2]/[(M_\sigma^2-4 M_\pi^2)s]}$;
\item the  masses of the
pseudoscalar mesons have been taken to be  $M_\pi=0.4$ GeV;
\item the external momenta are
$\vec  q_1 =(1,1,0)$ and $\vec  q_2=(-1,-1,0)$,
so that  there are  only two intermediate states, 
of zero total momentum, corresponding to $k=0$ and $k=2 \pi/L$,
 and energy less than 
the energy of the external state; of the two states, the one with
the two pions at rest does not contribute since $\delta(4 M_\pi^2)=0$. 
\item $A^W(s)$ has been taken to be constant and equal to $1$.
\end{enumerate}
By fitting the curve of fig. 1 as a function of $t_2$ it is possible to
extract $A(s_q) \cos(s_q)$, with $s_q=4 [M_\pi^2+2 (2 \pi/La)^2]$, and
$A(s_k) \sin(s_k)$, with $s_k= 4[M_\pi^2+ (2 \pi/La)^2]$. In the
same way, by increasing 
the energy of the external two-pion state, and using also states with
non-zero momentum, we can extract $A_c$ and $A_s$ as a function of $s_q$
and extrapolate the amplitude and the phase to their physical value.
\section*{Conclusion}
We have shown that, in spite of the MTNGT, it is possible to extract 
the relevant information on two-body decay matrix elements, under
the hypotheses that final-state interactions are dominated by nearby
resonances and that the couplings are smooth in the momenta.
We have also outlined  the strategy to extract 
this information from the Euclidean  correlation functions that
can be computed in standard numerical simulations. The feasibility
of our proposal requires a dedicated study, which will be presented 
elsewhere.
\section*{Acknowledgements}
We warmly thank M. Testa  for many useful discussions.
G.M.  thanks the Theory Division of CERN for the
 kind hospitality during the completion of this work.
We acknowledge the partial support by  M.U.R.S.T., Italy.

\newpage 
\begin{figure}[t]   
    \begin{center}
       \setlength{\unitlength}{1truecm}
       \begin{picture}(6.0,6.0)   
          \put(-5.2,-5.4){\includegraphics{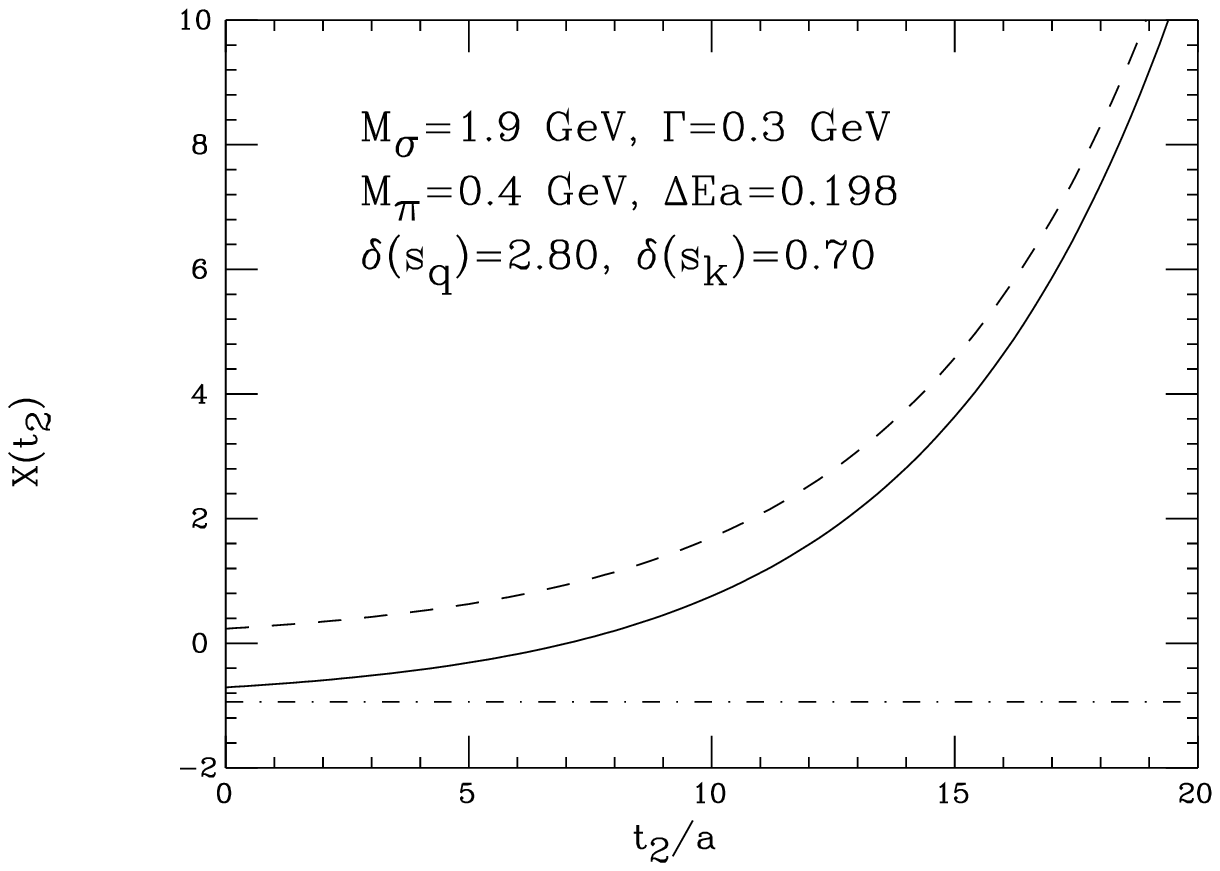}}  
       \end{picture}
    \end{center}
    \caption[]{\it{$X(t_2)=
R^{{\cal H}_W}(t_2,\vec q)$ as a function of  $t_2$
with the parameters given  in the text. $\Delta E a=2(E_{\vec q}-E_
{\vec k})a$, where $2 E_{\vec q}=\sqrt{s_q}=2\sqrt{M_\pi^2+2 (2 \pi/La)^2}$ and
$2E_{\vec k}=\sqrt{s_k}=2\sqrt{M_\pi^2+ (2 \pi/La)^2}$.
$\delta(s_q)$ and $\delta(s_k)$ are the corresponding  phases.
The dot-dashed line is the constant term corresponding
to $A^W(s_q) \cos \delta(s_q)$, the dashed curve is the
contribution of $A^W(s_k) \sin \delta(s_k)$ and the solid curve is the sum of 
the two terms.}}    \protect\label{esempio}
\end{figure}
\end{document}